\documentclass[%
reprint,
 amsmath,amssymb,
 aps,
]{revtex4-2}
\usepackage{color}
\newcommand{\editR}[1]{{\color{black}{#1}}}
\usepackage[dvipsnames]{xcolor}
\usepackage[utf8]{inputenc}
\usepackage{wrapfig}
\usepackage[]{graphicx,xcolor}
\usepackage{tabularx}
\usepackage{booktabs}
\usepackage{textcomp}
\usepackage{amsmath}
\usepackage{amssymb}
\usepackage[]{graphics}
\usepackage{amssymb}
\usepackage{amsmath}
\usepackage{color}
\usepackage{ifpdf}
\usepackage{lipsum}
\usepackage{siunitx}
\usepackage{braket}
\usepackage{soul}
\usepackage{tikz}
\usetikzlibrary{automata,positioning}
\usepackage{dcolumn}% Align table columns on decimal point
\usepackage{bm}
\graphicspath{ {./pics/} }% пакет для задания полей страницы командой \geometry

\begin{document}
\title{Resonant mode coupling approximation for calculation of optical spectra of photonic crystal slabs}

\author{Dmitrii~A.~Gromyko}
\email[]{e-mail: Dmitrii.Gromyko@skoltech.ru}
\affiliation{Skolkovo Institute of Science and Technology, Nobel Street 3, 121205 Moscow, Russia}
%\affiliation{Lomonosov Moscow State University, Physics Department, Moscow}
% \affiliation{A.~M.~Prokhorov General Physics Institute, RAS, Vavilova 38, Moscow, Russia}

\author{Sergey~A.~Dyakov}
\affiliation{Skolkovo Institute of Science and Technology, Nobel Street 3, 121205 Moscow, Russia}

\author{Sergei~G.~Tikhodeev}
\affiliation{A.~M.~Prokhorov General Physics Institute, RAS, Vavilova 38, Moscow, Russia}
\affiliation{Faculty of Physics, Lomonosov Moscow State University, 119991 Moscow, Russia}

\author{Nikolay~A.~Gippius}
\affiliation{Skolkovo Institute of Science and Technology, Nobel Street 3, 121205 Moscow, Russia}

    \begin{abstract}
We develop the resonant mode coupling approximation to calculate the optical spectra of a stack of two photonic crystal slabs. The method is based on a derivation of the input and output resonant vectors in each slab in terms of the Fourier modal method in the scattering matrix form. We show that using the resonant mode coupling approximation of the scattering matrices of the upper and lower slabs, one can construct the total scattering matrix of the stack. 
The formation of the resonant output and input vectors of the stacked system is rigorously derived by means of an effective Hamiltonian. We demonstrate that the proposed procedure dramatically decreases the computation time without sufficient loss of accuracy. We believe that the proposed technique can be a powerful tool for fast solving inverse scattering problems using stochastic optimization methods such as genetic algorithms or machine learning. 
    \end{abstract}        
\maketitle

\section{Introduction}
The development of methods for fabricating nanostructures in the last three decades has led to the emergence of many theoretical tools for the simulation of their optical properties. A great variety of theoretical methods makes it possible to comprehensively study the optical resonances of the entire system and its parts, revealing the contribution of each structural element to the resulting optical response before proceeding with their manufacture \cite{Yee1966,Draine1994}. In the engineering of photonic structures and devices, we always deal with photonic modes that are essential for understanding the structures' physical properties. By changing the geometry and composition of a structure, we tune the modes' localization, spectral position, decay time and make them hybridize \cite{friedrich1985interfering,Petschulat2008,Liu2009,limonov2017fano,bogdanov2019bound}. Due to the great opportunities given by fabrication methods, one can vary many geometrical parameters in wide ranges to achieve the desired optical response, taking a long CPU time to perform. On the other side, in many practically important cases, the structure of interest consists of several parts or subsystems whose optical properties are already known. With this regard, one can consider, for example, a stack of two photonic crystal slabs like in Refs.~\cite{Gippius2010,Weiss2011a}. In these works, it has been demonstrated that one can determine the resonant frequencies of the stack using a resonant mode coupling approximation based on the resonant modes \cite{Tikhodeev2002b,Weiss2017,Bykov2013,Lalanne2019} of the photonic crystal slabs found within the formalism of the Fourier modal method \editR{(FMM)} \cite{whittaker1999scattering, Tikhodeev2002b} also known as rigorous coupled-wave analysis (RCWA) \cite{moharam1995formulation}. It allows fast and efficient variation of the stack's geometrical parameters, for instance, the lateral shifts of the slabs or their thicknesses. See in, e.g., Ref.~\cite{Kristensen2020} for other various applications of the concept of the resonant or quasinormal modes.

In this article, we present a further development of the resonant mode coupling approximation to calculate the stack's optical spectra based on the knowledge of the upper and lower photonic crystal slabs' eigenmodes. We show that once we know all the resonant energies and the input and output resonant vectors, we can immediately construct a resonant approximation for the scattering matrix of the whole system for any given lateral shift between the subsystems.
We construct an effective Hamiltonian of the stacked system that provides a clear understanding of the interaction of the resonant modes. After that, considering a numerical example and comparing the approximate reflectance spectra of the stack with the results of calculations made by the standard Fourier modal method, we show their good quantitative agreement.

\section{Coupled resonances excited by an external plane wave}
By definition, a scattering matrix is the matrix function that provides a relation between the incoming and outgoing electromagnetic waves. According to \cite{Gippius2005c}, any scattering matrix can be presented %in the form of 
using the resonant approximation:
\begin{equation}
    \left(\begin{array}{c}
\left|d_{2}\right\rangle \\
\left|u_{1}\right\rangle
\end{array}\right)=\left[\tilde{\mathrm{S}}+\sum_{n=1}^{N}\left|O_{n}\right\rangle \frac{1}{\omega-\omega_{n}}\left\langle I_{n}\right|\right]\left(\begin{array}{c}
\left|d_{1}\right\rangle \\
\left|u_{2}\right\rangle
\end{array}\right).
\label{ResApprox_Definition}
\end{equation}
Here $\left|d,u\right\rangle$ are the vectors consisting of the complex amplitudes of the electromagnetic waves propagating downwards and upwards respectively. Their subscript indices 1 and 2 indicate that these amplitudes are calculated infinitesimally above the upper boundary and below the lower boundary, respectively. The matrix $\tilde{\mathrm{S}}$ is a slowly varying matrix function of energy, which is called henceforth a background scattering matrix. All the sharp features of the optical spectrum are described by the resonant terms that are first order poles $\omega_n$ with residuals $\left|O_{n}\right\rangle \left\langle I_{n}\right|$. We refer to the vectors $\left|O_{n}\right\rangle$ and $\left\langle I_{n}\right|$ as output and input resonant vectors.

\begin{figure}
    \centering
    \includegraphics[width=1\linewidth]{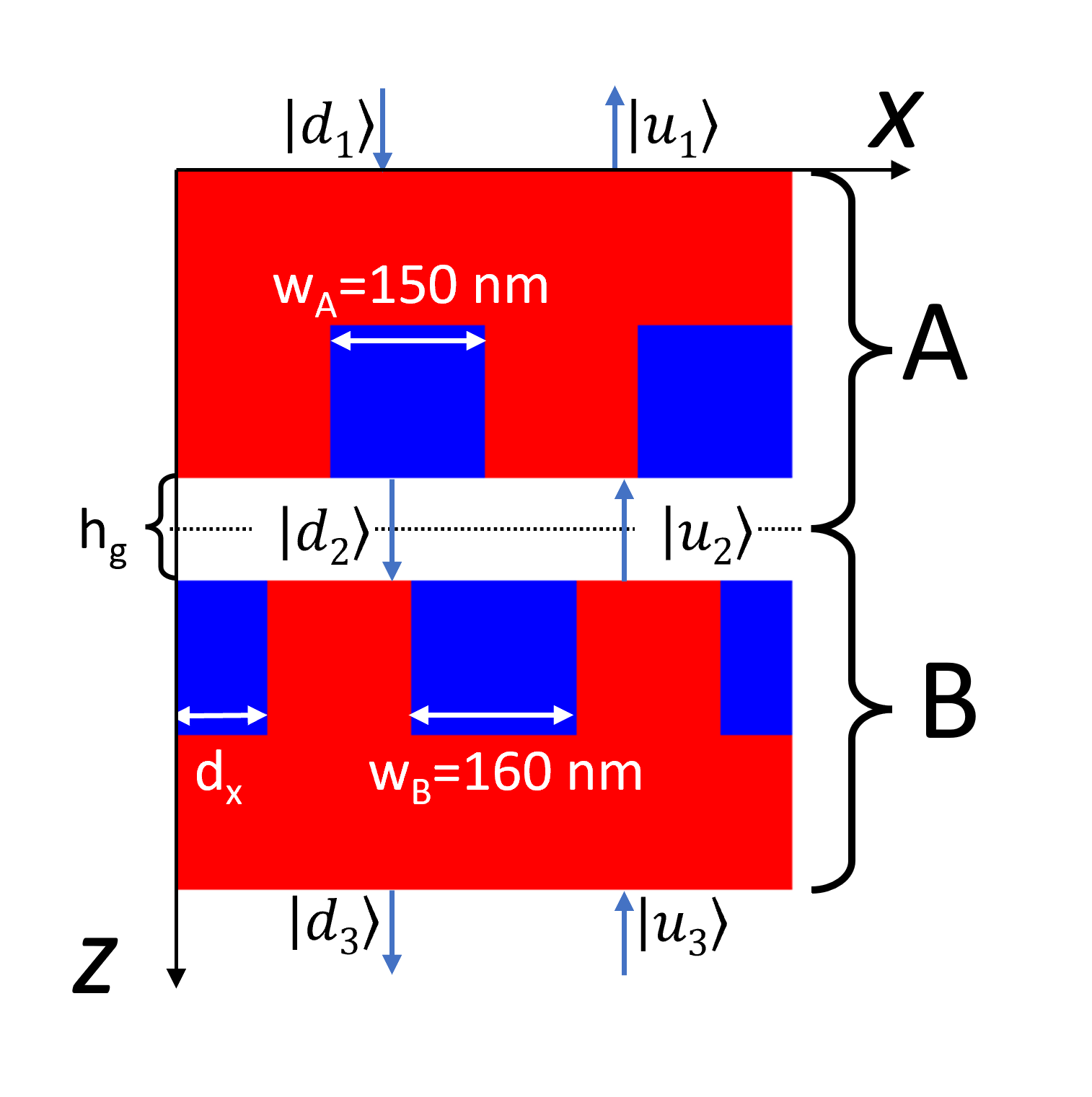}
    \caption{Schematic cross-section of two unit cells of the model structure. Red color denotes the crystalline silicon, blue color denotes SiO$_2$. The whole structure consists of two subsystems A and B separated by the air gap of the thickness $h_g$. Each subsystem consists of two layers (homogeneous and periodic) and three interfaces (two external interfaces with air and an inner one). All the layers are 150-nm thick, widths of the upper and lower periodic SiO$_2$ wires are $w_A=150$~nm and $w_B=160$~nm correspondingly. Lower subsystem B is shifted relative to upper system A  at a distance $d_x$, which is 90~nm for the particular configuration.}
    \label{fig:struct}
\end{figure}

Suppose we have a system that is composed of two subsystems A and B {(see in Fig.~\ref{fig:struct})}, {which scattering matrices are described using the resonant approximation with $N$ and $M$ resonances correspondingly}:
%In order to obtain the expressions for the input and output resonant vectors of the whole system, we consider a resonant approximation for the scattering matrices of the subsystems:
\begin{equation}
\begin{aligned}
&\hspace*{-0.2cm}\left(\begin{array}{c}
\hspace*{-0.13cm}\left|d_{2}\right\rangle \\
\hspace*{-0.13cm}\left|u_{1}\right\rangle
\end{array}\hspace*{-0.13cm}\right)\hspace*{-0.13cm}=\hspace*{-0.13cm}\left[\left(\hspace*{-0.13cm}\begin{array}{cc}
\tilde{\mathrm{S}}_{d d}^{\mathrm{a}} & \tilde{\mathrm{S}}_{du}^{\mathrm{a}} \\
\tilde{\mathrm{S}}_{u d}^{\mathrm{a}} & \tilde{\mathrm{S}}_{u u}^{\mathrm{a}}
\end{array}\hspace*{-0.13cm}\right)\hspace*{-0.13cm}+\hspace*{-0.13cm}\sum_{n=1}^{N}\left|O_{n}^{\mathrm{a}}\right\rangle \frac{1}{\omega-\omega_{n}^{a}}\left\langle I_{n}^{\mathrm{a}}\right|\right]\hspace*{-0.13cm}\left(\begin{array}{c}
\hspace*{-0.13cm}\left|d_{1}\right\rangle \\
\hspace*{-0.13cm}\left|u_{2}\right\rangle
\end{array}\hspace*{-0.13cm}\right),\\
&\hspace*{-0.2cm}\left(\hspace*{-0.13cm}\begin{array}{c}
\left|d_{3}\right\rangle \\
\left|u_{2}\right\rangle
\end{array}\hspace*{-0.13cm}\right)\hspace*{-0.13cm}=\hspace*{-0.13cm}\left[\left(\hspace*{-0.13cm}\begin{array}{cc}
\tilde{\mathrm{S}}_{d d}^{\mathrm{b}} & \tilde{\mathrm{S}}_{d u}^{\mathrm{b}} \\
\tilde{\mathrm{S}}_{u d}^{\mathrm{b}} & \tilde{\mathrm{S}}_{u u}^{\mathrm{b}}
\end{array}\hspace*{-0.13cm}\right)\hspace*{-0.13cm}+\hspace*{-0.13cm}\sum_{n=1}^{M}\left|O_{n}^{\mathrm{b}}\right\rangle \frac{1}{\omega-\omega_{n}^{\mathrm{b}}}\left\langle I_{n}^{\mathrm{b}}\right|\right]\hspace*{-0.13cm}\left(\hspace*{-0.13cm}\begin{array}{c}
\left|d_{2}\right\rangle \\
\left|u_{3}\right\rangle
\end{array}\hspace*{-0.13cm}\right).
\end{aligned}
\label{resonancesAB}
\end{equation}
Here subscript indices 1, 2, and 3 correspond to the waves propagating above the structure A, between structures A and B, and below structure B respectively. The procedure for calculating the resonant energies of the stacked system was presented in \cite{Gippius2010}. In what follows we derive the relation between amplitudes $\left|d_{3}, u_1\right\rangle$ and $\left|d_{1}, u_3\right\rangle$ in form \eqref{ResApprox_Definition}. It is essential to underline that we are looking for analytical expressions for the background matrix, the input and output vectors. Direct combination of the scattering matrices calculated for each energy using \eqref{resonancesAB} does not provide us any knowledge of the stacked system eigenmodes and does not ensure any significant computational speed up.

% The resonant input vectors of the stacked system should be dependent on the type of excitation. For example, this could be an external plane wave or an oscillating current located between the subsystems A and B. For the case of external plane waves we write the resonant approximation in the desired form:
% \begin{equation}
% \left(\hspace*{-0.13cm}\begin{array}{c}
% \left|d_{3}\right\rangle \\
% \left|u_{1}\right\rangle
% \end{array}\hspace*{-0.13cm}\right)\hspace*{-0.13cm}=\hspace*{-0.13cm}\left[\left(\hspace*{-0.13cm}\begin{array}{cc}
% \tilde{\mathrm{S}}_{d d} & \tilde{\mathrm{S}}_{du} \\
% \tilde{\mathrm{S}}_{u d} & \tilde{\mathrm{S}}_{u u}
% \end{array}\hspace*{-0.13cm}\right)\right.\hspace*{-0.1cm}+
% \hspace*{-0.3cm}\left.\sum_{i=1}^{L=N+M}\hspace*{-0.35cm}\left|O_{i}\right\rangle \frac{1}{\omega-\omega_{i}}\left\langle I_{i}\right|\right]\hspace*{-0.13cm}\left(\hspace*{-0.13cm}\begin{array}{c}
% \left|d_{1}\right\rangle \\
% \left|u_{3}\right\rangle
% \end{array}\hspace*{-0.13cm}\right).
% \label{Desired}
% \end{equation}
% \begin{equation}
%  \mathrm{S}^{0}=\left(\begin{array}{cc}
% \tilde{\mathrm{S}}_{d d}^{\mathrm{0}} & \tilde{\mathrm{S}}_{d u}^{\mathrm{0}} \\
% \tilde{\mathrm{S}}_{u d}^{\mathrm{0}} & \tilde{\mathrm{S}}_{u u}^{\mathrm{0}}
% \end{array}\right)+\sum_{i=1}^{L}\left|Out_{i}\right\rangle \frac{1}{\omega-\omega_{i}}\left\langle In_{i}\right|
% \end{equation}
A resonant mode is a non-trivial solution of the Maxwell's equations without sources \cite{Gippius2005c,Muljarov2010,Lalanne2019}. Following this definition, one should assume that there are no incoming waves in the system, i.e. $\left|d_{1}\right\rangle=0$,  $\left|u_{3}\right\rangle=0$. In previous works \cite{Gippius2010, Weiss2011a} the coefficients of the resonant excitation were defined as a product of the resonant input vectors and the only non-zero wave amplitudes $\left|d_2,u_{2}\right\rangle$.
Now, in order to derive the resonant mode coupling approximation for the scattering matrix, we use a similar approach.
However, for our problem of calculating the optical spectra, the presence of incoming waves $\left|d_1,\,u_3\right\rangle\ne 0$ is essential. We define the resonant coefficients, so that they also account for the incoming waves:
\begin{equation}
\begin{array}{l}
    (\omega-\omega_{n}^{a})\alpha_{n}=
    \left\langle I_{n}^{\mathrm{a}} \right|
    \begin{pmatrix}
    \left|d_{1}\right\rangle\\
    \left|u_{2}\right\rangle
    \end{pmatrix}\hspace*{-0.1cm}=\hspace*{-0.1cm}\left\langle I_{u, n}^{\mathrm{a}}|u_2\right\rangle\hspace*{-0.07cm}+\hspace*{-0.07cm}\left\langle I_{d, n}^{\mathrm{a}}|d_1\right\rangle,\\
        (\omega-\omega_{n}^{b})\beta_{n}=
    \left\langle I_{n}^{\mathrm{b}} \right|
    \begin{pmatrix}
    \left|d_{2}\right\rangle\\
    \left|u_{3}\right\rangle
    \end{pmatrix}\hspace*{-0.1cm}=\hspace*{-0.1cm}\left\langle I_{d, n}^{\mathrm{b}}|d_2\right\rangle\hspace*{-0.07cm}+\hspace*{-0.07cm}\left\langle I_{u, n}^{\mathrm{b}}|u_3\right\rangle.
    \end{array}
    \label{coefs_definition}
\end{equation}

These coefficients show how the incoming electromagnetic waves excite the resonances of the subsystems:
\begin{equation}
\begin{aligned}
&\left(\begin{array}{c}
\left|d_{2}\right\rangle \\
\left|u_{1}\right\rangle
\end{array}\right)=\left(\begin{array}{cc}
\tilde{\mathrm{S}}_{d d}^{\mathrm{a}} & \tilde{\mathrm{S}}_{du}^{\mathrm{a}} \\
\tilde{\mathrm{S}}_{u d}^{\mathrm{a}} & \tilde{\mathrm{S}}_{u u}^{\mathrm{a}}
\end{array}\right)\left(\begin{array}{c}
\left|d_{1}\right\rangle \\
\left|u_{2}\right\rangle
\end{array}\right)+\sum_{n=1}^{N}\left|O_{n}^{\mathrm{a}}\right\rangle \alpha_n,\\
&\left(\begin{array}{c}
\left|d_{3}\right\rangle \\
\left|u_{2}\right\rangle
\end{array}\right)=\left(\begin{array}{cc}
\tilde{\mathrm{S}}_{d d}^{\mathrm{b}} & \tilde{\mathrm{S}}_{d u}^{\mathrm{b}} \\
\tilde{\mathrm{S}}_{u d}^{\mathrm{b}} & \tilde{\mathrm{S}}_{u u}^{\mathrm{b}}
\end{array}\right) \left(\begin{array}{c}
\left|d_{2}\right\rangle \\
\left|u_{3}\right\rangle
\end{array}\right)+\sum_{n=1}^{M}\left|O_{n}^{\mathrm{b}}\right\rangle \beta_n.
\end{aligned}
\label{resonancesAB_with_coefs}
\end{equation}

\editR{Denoting the identity matrix as $\mathbb{I}$} we extract vectors $\left|d_{2},u_2\right\rangle$ from equations (\ref{resonancesAB_with_coefs}):
\begin{gather}
    \label{eq_d2}
\mathbb{D}_{dd}^{-1}\left|d_2\right\rangle\equiv(\editR{\mathbb{I}}-\tilde{\mathrm{S}}_{d u}^{\mathrm{a}}\tilde{\mathrm{S}}_{u d}^{\mathrm{b}})\left|d_2\right\rangle=\\
    \sum_{n=1}^N\left|O_n^{\mathrm{a}}\right\rangle\alpha_n+\sum_{n=1}^M\tilde{\mathrm{S}}_{d u}^{\mathrm{a}}\left|O_{u,n}^{\mathrm{b}}\right\rangle\beta_n +
    \tilde{\mathrm{S}}_{d d}^{\mathrm{a}}\left| d_1\right\rangle+\tilde{\mathrm{S}}_{d u}^{\mathrm{a}}\tilde{\mathrm{S}}_{u u}^{\mathrm{b}}\left|u_3\right\rangle, \nonumber
    \end{gather}
    \begin{gather}
    \label{eq_u2}
    \mathbb{D}_{uu}^{-1}\left|u_2\right\rangle\equiv(\editR{\mathbb{I}}-\tilde{\mathrm{S}}_{u d}^{\mathrm{b}}\tilde{\mathrm{S}}_{d u}^{\mathrm{a}})\left|u_2\right\rangle=\\
    \sum_{n=1}^N\tilde{\mathrm{S}}_{u d}^{\mathrm{b}}\left|O_n^{\mathrm{a}}\right\rangle\alpha_n+\sum_{n=1}^M\left|O_{u,n}^{\mathrm{b}}\right\rangle\beta_n +
    \tilde{\mathrm{S}}_{u d}^{\mathrm{b}}\tilde{\mathrm{S}}_{d d}^{\mathrm{a}}\left| d_1\right\rangle+\tilde{\mathrm{S}}_{u u}^{\mathrm{b}}\left|u_3\right\rangle. \nonumber
\end{gather}
\editR{Vectors $\left|d_{2},u_2\right\rangle$ are amplitudes of the Fourier harmonics that propagate between two subsystems. One can use them to calculate the electric and magnetic fields distributions inside the homogeneous intermediate layer.}
Now we obtain the new system that defines the resonant coefficients by substituting \eqref{eq_d2}, \eqref{eq_u2} into \eqref{coefs_definition} and using the matrix relation $(\mathbb{I}-\mathrm{AB})^{-1}\mathrm{A}=\mathrm{A}(\mathbb{I}-\mathrm{BA})^{-1}$:
\begin{gather}
\left(\omega-\Omega^{\mathrm{a}}\right) \alpha=\left\langle I_{u}^{\mathrm{a}} \right|\tilde{\mathrm{S}}_{ud}^{\mathrm{b}} \mathbb{D}_{dd}\left|{O}_{d}^{\mathrm{a}}\right\rangle \alpha+\left\langle I_{u}^{\mathrm{a}}|\mathbb{D}_{uu}| {O}_{u}^{\mathrm{b}}\right\rangle \beta+\nonumber\\
+\left\langle I_{d}^{\mathrm{a}}\right.\left|d_1\right\rangle+
\left\langle I_{u}^{\mathrm{a}}\right|\tilde{\mathrm{S}}_{ud}^{\mathrm{b}}\mathbb{D}_{dd}\tilde{\mathrm{S}}_{dd}^{\mathrm{a}}\left|d_1\right\rangle
+\left\langle I_{u}^{\mathrm{a}}\right|\mathbb{D}_{uu}\tilde{\mathrm{S}}_{uu}^{\mathrm{b}}\left|u_3\right\rangle,
\label{system_on_alpha_beta1}
\\
\left(\omega -\Omega^{\mathrm{b}}\right) \beta=
\left\langle I_{d}^{\mathrm{b}}| \mathbb{D}_{dd}| {O}_{d}^{\mathrm{a}}\right\rangle \alpha+
\left\langle I_{d}^{\mathrm{b}}\left|\tilde{\mathrm{S}}_{du}^{\mathrm{a}}\mathbb{D}_{uu}\right| {O}_{u}^{\mathrm{b}}\right\rangle \beta+\nonumber\\
\hspace*{-0.15cm}+\left\langle I_{d}^{\mathrm{b}}\right|\mathbb{D}_{dd}\tilde{\mathrm{S}}_{dd}^{\mathrm{a}}\left|d_1\right\rangle+
\left\langle I_{u}^{\mathrm{b}}\right.\left|u_3\right\rangle+
\left\langle I_{d}^{\mathrm{b}}\right|\tilde{\mathrm{S}}_{du}^{\mathrm{a}}\mathbb{D}_{uu}\tilde{\mathrm{S}}_{uu}^{\mathrm{b}}\left|u_3\right\rangle, \label{system_on_alpha_beta2}
\\
\Omega^{\mathrm{a,b}}=\text{diag}\{\omega^{\mathrm{a,b}}_n\}.\nonumber
\end{gather}
\editR{Note that matrices $\mathbb{D}_{uu}, \mathbb{D}_{dd}$ are Fabry-Perot operators that describe  multiple internal reflections between the subsystems.}

% Here we use the following definition of the "dressed" resonant output vector, which is a sum of an initial output vector and all its non-resonant reflections:
% \begin{equation}
% \begin{aligned}
% &\left|\mathbf{O}_{d, n}^{\mathrm{a}}\right\rangle=\underbrace{\left(\mathbb{I} -\tilde{\mathrm{S}}_{d u}^{\mathrm{a}} \tilde{\mathrm{S}}_{u d}^{\mathrm{b}}\right)^{-1}}_{\mathbb{D}_1}\left|O_{d, n}^{\mathrm{a}}\right\rangle, \\
% %%
% &\left|\mathbf{O}_{u, n}^{\mathrm{b}}\right\rangle=\underbrace{\left(\mathbb{I} -\tilde{\mathrm{S}}_{u d}^{\mathrm{b}} \tilde{\mathrm{S}}_{d u}^{\mathrm{a}}\right)^{-1}}_{\mathbb{D}_2}\left|O_{u, n}^{\mathrm{b}}\right\rangle,\\
% &\qquad\quad\Omega^{\mathrm{a,b}}=\text{diag}\{\omega^{\mathrm{a,b}}\}.\\
% \label{Dressed}
% \end{aligned}
% \end{equation}

One can rewrite system \eqref{system_on_alpha_beta1}, \eqref{system_on_alpha_beta2} in the following form so that the connection with the eigenproblem on resonant states of the stacked system would be obvious:
\begin{gather}
\label{Eigen_EM}
\omega\left(\begin{array}{c}
\alpha \\
\beta
\end{array}\right)={\underbrace{\left(\begin{array}{cc}
\mathrm{H}_{aa} &\mathrm{H}_{ab} \\
\mathrm{H}_{ba} & \mathrm{H}_{bb}
\end{array}\right)}_H}
\left(\begin{array}{c}
\alpha \\
\beta
\end{array}\right)+\left\langle I\right|\left(\begin{array}{c}
\left|d_1\right\rangle\\
\left|u_3\right\rangle
\end{array}\right),\\
\left\langle I\right|=
\left(\hspace*{-0.13cm}\begin{array}{cc}
\left\langle I_{d}^{\mathrm{a}}\right|+\left\langle I_{u}^{\mathrm{a}}\right|\tilde{\mathrm{S}}_{ud}^{\mathrm{b}}\mathbb{D}_{dd}\tilde{\mathrm{S}}_{dd}^{\mathrm{a}}& \hspace*{-0.3cm}\left\langle I_{u}^{\mathrm{a}}\right|\mathbb{D}_{uu}\tilde{\mathrm{S}}_{uu}^{\mathrm{b}} \\
\left\langle I_{d}^{\mathrm{b}}\right|\mathbb{D}_{dd}\tilde{\mathrm{S}}_{dd}^{\mathrm{a}} & 
\hspace*{-0.3cm}\left\langle I_{u}^{\mathrm{b}}\right|+\left\langle I_{d}^{\mathrm{b}}\right|\tilde{\mathrm{S}}_{du}^{\mathrm{a}}\mathbb{D}_{uu}\tilde{\mathrm{S}}_{uu}^{\mathrm{b}}
\end{array}\hspace*{-0.13cm}\right).\nonumber
\end{gather}
Here $H$ is an effective Hamiltonian
\begin{equation}
    \begin{pmatrix}
    \Omega^a\hspace*{-0.1cm}+\hspace*{-0.1cm}\left\langle I_{u}^{\mathrm{a}}\left|\tilde{\mathrm{S}}_{u d}^{\mathrm{b}}\mathbb{D}_{dd}\right| {O}_{d}^{\mathrm{a}}\right\rangle &
    \hspace*{-0.5cm}\left\langle I_{u}^{\mathrm{a}}| \mathbb{D}_{uu}| {O}_{u}^{\mathrm{b}}\right\rangle\\
    \hspace*{-0.1cm}\left\langle I_{d}^{\mathrm{b}}| \mathbb{D}_{dd}| {O}_{d}^{\mathrm{a}}\right\rangle &
    \hspace*{-0.6cm}\Omega^b\hspace*{-0.1cm}+\hspace*{-0.1cm}\left\langle I_{d}^{\mathrm{b}}\left|\tilde{\mathrm{S}}_{d u}^{\mathrm{a}}\mathbb{D}_{uu}\right| {O}_{u}^{\mathrm{b}}\right\rangle
    \hspace*{-0.07cm}
    \end{pmatrix}
    \hspace*{-0.07cm}\equiv\hspace*{-0.07cm}{\begin{pmatrix}
    \hspace*{-0.07cm} H_{{aa}} & \hspace*{-0.1cm}H_{{ab}}\\
    \hspace*{-0.07cm} H_{{ba}} & \hspace*{-0.1cm}H_{{bb}}
    \hspace*{-0.07cm} \end{pmatrix}},
    \label{Eigensystem}
\end{equation}
whose eigenvalues $\omega^{\mathrm{c}}_n$ are the resonant energies of the stacked system \cite{Gippius2010} and the corresponding eigenvectors are the resonant coefficients that define the partial contributions of each resonance of subsystems A and B into the resonances of the stacked system.
The second term in the right part of equation~\eqref{Eigen_EM} is a total force that excites the resonances of the stacked system. Note that eigenvalues and eigenvectors of matrix $H$ are non-trivial solutions of equation~\eqref{Eigen_EM} with no external excitation $\left|d_1,\, u_3\right\rangle=0$ and thus represent the resonant modes.

To solve system of equations \eqref{Eigen_EM} one has to find all eigenvalues and eigenvectors of the Hamiltonian \eqref{Eigensystem}:
\begin{equation}
    HX=X\Omega^c,\quad  \Omega^{\mathrm{c}}=\text{diag}\{\omega^{\mathrm{c}}_n\}.
\end{equation}
Here $\Omega^{\mathrm{c}}$ is a diagonal matrix of the resonant energies of the stacked system $\omega^{\mathrm{c}}_n$, $X$ is a square matrix whose columns are the corresponding right eigenvectors $[\alpha^{(n)},\beta^{(n)}]^T$ of the Hamiltonian. 

Taking the inverse operator we obtain:
\begin{equation}
    \begin{pmatrix}
    \alpha\\
    \beta
    \end{pmatrix}=X\dfrac{1}{\omega \editR{\mathbb{I}}-\Omega^{\mathrm{c}}}X^{-1}\left\langle I\right|\left(\begin{array}{c}
\left|d_1\right\rangle\\
\left|u_3\right\rangle
\end{array}\right).
\end{equation}
%here \editR{$\mathbb{I}$} is the identity matrix of size $(N+M)\times(N+M)$.

% \begin{equation}
% \begin{aligned}
% A\omega=HA+f\\
% P^{-1}HP=H_d\\
% eig(H)=[P,H_d]\\
% H=PH_dP^{-1}\\
% P^{-1}A\omega=H_dP^{-1}A+P^{-1}F\\
% H_d=\mathrm{diag}\{\omega_i\}\\
% A=P\dfrac{1}{\omega-\omega_i}P^{-1}\mathrm{Imat}\left(\begin{array}{c}
% \left|d_1\right\rangle\\
% \left|u_3\right\rangle
% \end{array}\right)
% \end{aligned}
% \end{equation}
% \begin{equation}
% \left\langle In\right|=P^{-1}\mathrm{Imat}
% \end{equation}
{With the given coefficients we can compute $\left|d_{2},u_2\right\rangle$ using Eq.~\eqref{eq_d2},\eqref{eq_u2} and substitute the result into \eqref{resonancesAB_with_coefs} in order to finally derive the outgoing amplitudes $\left|d_{3},u_1\right\rangle$:} 
\begin{widetext}
\begin{gather}
    \begin{pmatrix}
    \left|d_3\right\rangle\\
    \left|u_1\right\rangle
    \end{pmatrix}\hspace*{-0.07cm}=\hspace*{-0.07cm}\begin{pmatrix}
    \hspace*{-0.5cm}\tilde{\mathrm{S}}^{\mathrm{b}}_{dd}\mathbb{D}_{dd}\left|{O}_{d}^{\mathrm{a}}\right\rangle & \hspace*{-0.7cm} \left|O^{\mathrm{b}}_d\right\rangle+\tilde{\mathrm{S}}^{\mathrm{b}}_{dd}\tilde{\mathrm{S}}_{du}^{\mathrm{a}}\mathbb{D}_{uu}\left|{O}_{u}^{\mathrm{b}}\right\rangle\\
     \left|O_{u}^{\mathrm{a}}\right\rangle+\tilde{\mathrm{S}}_{u u}^{\mathrm{a}}\tilde{\mathrm{S}}_{u d}^{\mathrm{b}}\mathbb{D}_{dd}\left|{O}_{d}^{\mathrm{a}}\right\rangle & 
\hspace*{-0.5cm}\tilde{\mathrm{S}}_{u u}^{\mathrm{a}}\mathbb{D}_{uu}\left|{O}_{u}^{\mathrm{b}}\right\rangle
    \end{pmatrix}
    \hspace*{-0.07cm}
    \begin{pmatrix}
    \alpha\\
    \beta
    \end{pmatrix}
    \hspace*{-0.07cm}+\hspace*{-0.07cm}\begin{pmatrix}
    \hspace*{-0.5cm}\tilde{\mathrm{S}}^{\mathrm{b}}_{dd}\mathbb{D}_{dd}\tilde{\mathrm{S}}^{\mathrm{a}}_{dd} &  \hspace*{-0.7cm}\tilde{\mathrm{S}}^{\mathrm{b}}_{du}+\tilde{\mathrm{S}}^{\mathrm{b}}_{dd}\tilde{\mathrm{S}}^{\mathrm{a}}_{du}\mathbb{D}_{uu}\tilde{\mathrm{S}}^{\mathrm{b}}_{uu}\\
     \tilde{\mathrm{S}}_{ud}^{\mathrm{a}} + \tilde{\mathrm{S}}_{uu}^{\mathrm{a}}\tilde{\mathrm{S}}_{ud}^{\mathrm{b}}\mathbb{D}_{dd}\tilde{\mathrm{S}}_{dd}^{\mathrm{a}}&
         \hspace*{-0.3cm}\tilde{\mathrm{S}}_{uu}^{\mathrm{a}}\mathbb{D}_{uu}\tilde{\mathrm{S}}_{uu}^{\mathrm{b}}
    \end{pmatrix}
    \hspace*{-0.07cm}\begin{pmatrix}
    \left|d_1\right\rangle\\
    \left|u_3\right\rangle
    \end{pmatrix}
    %%
    %%
%     \hspace*{0.6cm}\left|u_1\right\rangle=\begin{pmatrix}
%     \left|O_{u}^{\mathrm{a}}\right\rangle+\tilde{\mathrm{S}}_{u u}^{\mathrm{a}}\tilde{\mathrm{S}}_{u d}^{\mathrm{b}}\mathbb{D}_{dd}\left|{O}_{d}^{\mathrm{a}}\right\rangle & 
%  \quad\tilde{\mathrm{S}}_{u u}^{\mathrm{a}}\mathbb{D}_{uu}\left|{O}_{u}^{\mathrm{b}}\right\rangle
%     \end{pmatrix}\begin{pmatrix}
%     \alpha\\
%     \beta
%     \end{pmatrix}
%     +\begin{pmatrix}
%      \tilde{\mathrm{S}}_{ud}^{\mathrm{a}} + \tilde{\mathrm{S}}_{uu}^{\mathrm{a}}\tilde{\mathrm{S}}_{ud}^{\mathrm{b}}\mathbb{D}_{dd}\tilde{\mathrm{S}}_{dd}^{\mathrm{a}}&
%          \quad\tilde{\mathrm{S}}_{uu}^{\mathrm{a}}\mathbb{D}_{uu}\tilde{\mathrm{S}}_{uu}^{\mathrm{b}}
%     \end{pmatrix}\begin{pmatrix}
%     \left|d_1\right\rangle\\
%     \left|u_3\right\rangle
%     \end{pmatrix}
\end{gather}
Now we can write the output vectors, input vectors, as well as the background matrix of the resonant approximation:
\begin{equation}
\left(\begin{array}{c}
\left|d_{3}\right\rangle \\
\left|u_{1}\right\rangle
\end{array}\right)=\left[\left(\begin{array}{cc}
\tilde{\mathrm{S}}_{d d}^{\mathrm{c}} & \tilde{\mathrm{S}}_{du}^{\mathrm{c}} \\
\tilde{\mathrm{S}}_{u d}^{\mathrm{c}} & \tilde{\mathrm{S}}_{u u}^{\mathrm{c}}
\end{array}\right)+
\sum_{n=1}^{L=N+M}\left|O^{\mathrm{c}}_{n}\right\rangle \frac{1}{\omega-\omega^c_{n}}\left\langle I^{\mathrm{c}}_{n}\right|\right]\left(\begin{array}{c}
\left|d_{1}\right\rangle \\
\left|u_{3}\right\rangle
\end{array}\right).
\label{Desired}
%%%%%
\end{equation}
\begin{equation}
\begin{aligned}
\left|O^{\mathrm{c}}\right\rangle=\begin{pmatrix}
\left|O^{\mathrm{c}}_1\right\rangle,& \left|O^{\mathrm{c}}_2\right\rangle,& ...
\end{pmatrix}=
\left[\begin{pmatrix}
\editR{\mathbb{O}} & \left|O^{\mathrm{b}}_d\right\rangle\\
\left|O^{\mathrm{a}}_u\right\rangle & \editR{\mathbb{O}}
\end{pmatrix} +
\begin{pmatrix}
\tilde{\mathrm{S}}^{\mathrm{b}}_{dd} & \editR{\mathbb{O}}\\
\editR{\mathbb{O}} & \tilde{\mathrm{S}}^{\mathrm{a}}_{uu}
\end{pmatrix} 
% \begin{pmatrix}
% \mathbb{D}_{dd} & \tilde{\mathrm{S}}_{du}^{\mathrm{a}}\mathbb{D}_{uu}\\
% \tilde{\mathrm{S}}_{ud}^{\mathrm{b}}\mathbb{D}_{dd} & \mathbb{D}_{uu}
% \end{pmatrix}
%
\begin{pmatrix}
\editR{\mathbb{I}} & \tilde{\mathrm{S}}_{du}^{\mathrm{a}}\\
\tilde{\mathrm{S}}_{ud}^{\mathrm{b}} & \editR{\mathbb{I}}
\end{pmatrix}
\begin{pmatrix}
\mathbb{D}_{dd} & \editR{\mathbb{O}}\\
\editR{\mathbb{O}} & \mathbb{D}_{uu}
\end{pmatrix}
\begin{pmatrix}
\left|O^{\mathrm{a}}_d\right\rangle & \editR{\mathbb{O}}\\
\editR{\mathbb{O}} & \left|O^{\mathrm{b}}_u\right\rangle
\end{pmatrix}\right]X,\\
%%
%%
% \editR{
% \left|O^{\mathrm{c}}\right\rangle=\begin{pmatrix}
% \left|O^{\mathrm{c}}_1\right\rangle,& \left|O^{\mathrm{c}}_2\right\rangle,& ...
% \end{pmatrix}=
% \left[\begin{pmatrix}
% \left|O^{\mathrm{a}}_u\right\rangle & 0\\
% 0 & \left|O^{\mathrm{b}}_d\right\rangle
% \end{pmatrix} +
% \begin{pmatrix}
% \tilde{\mathrm{S}}^{\mathrm{a}}_{uu} & 0\\
% 0 & \tilde{\mathrm{S}}^{\mathrm{b}}_{dd}
% \end{pmatrix} 
% % \begin{pmatrix}
% % \tilde{\mathrm{S}}_{ud}^{\mathrm{b}}\mathbb{D}_{dd} & \mathbb{D}_{uu}\\
% % \mathbb{D}_{dd} & \tilde{\mathrm{S}}_{du}^{\mathrm{a}}\mathbb{D}_{uu}
% % \end{pmatrix}
% \begin{pmatrix}
% \tilde{\mathrm{S}}_{ud}^{\mathrm{b}} & I\\
% I & \tilde{\mathrm{S}}_{du}^{\mathrm{a}}
% \end{pmatrix}
% \begin{pmatrix}
% \mathbb{D}_{dd} & 0\\
% 0 & \mathbb{D}_{uu}
% \end{pmatrix}
% \begin{pmatrix}
% \left|O^{\mathrm{a}}_d\right\rangle & 0\\
% 0 & \left|O^{\mathrm{b}}_u\right\rangle
% \end{pmatrix}\right]X,}\\
%%
%%
%%
% \left(\begin{array}{cc}
% \tilde{\mathrm{S}}^{\mathrm{b}}_{dd}\mathbb{D}_{dd}\left|{O}_{d}^{\mathrm{a}}\right\rangle &  \quad\left|O^{\mathrm{b}}_d\right\rangle+\tilde{\mathrm{S}}^{\mathrm{b}}_{dd}\tilde{\mathrm{S}}_{du}^{\mathrm{a}}\mathbb{D}_{uu}\left|{O}_{u}^{\mathrm{b}}\right\rangle \\
% \left|O_{u}^{\mathrm{a}}\right\rangle+\tilde{\mathrm{S}}_{u u}^{\mathrm{a}}\tilde{\mathrm{S}}_{u d}^{\mathrm{b}}\mathbb{D}_{dd}\left|{O}_{d}^{\mathrm{a}}\right\rangle & 
%  \quad\tilde{\mathrm{S}}_{u u}^{\mathrm{a}}\mathbb{D}_{uu}\left|{O}_{u}^{\mathrm{b}}\right\rangle
% \end{array}\right)X,\\
\left\langle I^{\mathrm{c}}\right|=\begin{pmatrix}
\left\langle I^{\mathrm{c}}_1\right|\\
\left\langle I^{\mathrm{c}}_2\right|\\
\vdots
\end{pmatrix}=
X^{-1}\left[
\begin{pmatrix}
\left\langle I_{d}^{\mathrm{a}}\right| & \editR{\mathbb{O}} \\
\editR{\mathbb{O}} & \left\langle I_{u}^{\mathrm{b}}\right|
\end{pmatrix}+
\begin{pmatrix}
\left\langle I_{u}^{\mathrm{a}}\right| & \editR{\mathbb{O}}\\
\editR{\mathbb{O}} & \left\langle I_{d}^{\mathrm{b}}\right|
\end{pmatrix}
% \begin{pmatrix}
% \tilde{\mathrm{S}}_{ud}^{\mathrm{b}}\mathbb{D}_{dd} & \mathbb{D}_{uu}\\
% \mathbb{D}_{dd} & \tilde{\mathrm{S}}_{du}^{\mathrm{a}}\mathbb{D}_{uu}
% \end{pmatrix}
\begin{pmatrix}
\tilde{\mathrm{S}}_{ud}^{\mathrm{b}} & \editR{\mathbb{I}}\\
\editR{\mathbb{I}} & \tilde{\mathrm{S}}_{du}^{\mathrm{a}}
\end{pmatrix}
\begin{pmatrix}
\mathbb{D}_{dd} & \editR{\mathbb{O}}\\
\editR{\mathbb{O}} & \mathbb{D}_{uu}
\end{pmatrix}
\begin{pmatrix}
\tilde{\mathrm{S}}_{dd}^{\mathrm{a}} & \editR{\mathbb{O}}\\
\editR{\mathbb{O}} & \tilde{\mathrm{S}}_{uu}^{\mathrm{b}}
\end{pmatrix}
\right],
% X^{-1}\left(\begin{array}{cc}
% \left\langle I_{d}^{\mathrm{a}}\right|+\left\langle I_{u}^{\mathrm{a}}\right|\tilde{\mathrm{S}}_{ud}^{\mathrm{b}}\mathbb{D}_{dd}\tilde{\mathrm{S}}_{dd}^{\mathrm{a}}& \left\langle I_{u}^{\mathrm{a}}\right|\mathbb{D}_{uu}\tilde{\mathrm{S}}_{uu}^{\mathrm{b}} \\
% \left\langle I_{d}^{\mathrm{b}}\right|\mathbb{D}_{dd}\tilde{\mathrm{S}}_{dd}^{\mathrm{a}} & \left\langle I_{u}^{\mathrm{b}}\right|+\left\langle I_{d}^{\mathrm{b}}\right|\tilde{\mathrm{S}}_{du}^{\mathrm{a}}\mathbb{D}_{uu}\tilde{\mathrm{S}}_{uu}^{\mathrm{b}}
% \end{array}\right),
\end{aligned}
\label{inputOut}
\end{equation}
\begin{equation}
    \tilde{\mathrm{S}}^{\mathrm{c}}=
    \begin{pmatrix}
    \editR{\mathbb{O}} & \tilde{\mathrm{S}}^{\mathrm{b}}_{du}\\
    \tilde{\mathrm{S}}^{\mathrm{a}}_{ud} & \editR{\mathbb{O}}
    \end{pmatrix}+
    \begin{pmatrix}
\tilde{\mathrm{S}}^{\mathrm{b}}_{dd} & \editR{\mathbb{O}}\\
\editR{\mathbb{O}} & \tilde{\mathrm{S}}^{\mathrm{a}}_{uu}
\end{pmatrix} 
% \begin{pmatrix}
% \mathbb{D}_{dd} & \tilde{\mathrm{S}}_{du}^{\mathrm{a}}\mathbb{D}_{uu}\\
% \tilde{\mathrm{S}}_{ud}^{\mathrm{b}}\mathbb{D}_{dd} & \mathbb{D}_{uu}
% \end{pmatrix}
\begin{pmatrix}
\editR{\mathbb{I}} & \tilde{\mathrm{S}}_{du}^{\mathrm{a}}\\
\tilde{\mathrm{S}}_{ud}^{\mathrm{b}} & \editR{\mathbb{I}}
\end{pmatrix}
\begin{pmatrix}
\mathbb{D}_{dd} & \editR{\mathbb{O}}\\
\editR{\mathbb{O}} & \mathbb{D}_{uu}
\end{pmatrix}
\begin{pmatrix}
\tilde{\mathrm{S}}_{dd}^{\mathrm{a}} & \editR{\mathbb{O}}\\
\editR{\mathbb{O}} & \tilde{\mathrm{S}}_{uu}^{\mathrm{b}}
\end{pmatrix},
%     \begin{pmatrix} \tilde{\mathrm{S}}_{dd}^{\mathrm{b}}\mathbb{D}_{dd}\tilde{\mathrm{S}}_{dd}^{\mathrm{a}}&
%   \tilde{\mathrm{S}}_{du}^{\mathrm{b}} + \tilde{\mathrm{S}}_{dd}^{\mathrm{b}}\tilde{\mathrm{S}}_{du}^{\mathrm{a}}\mathbb{D}_{uu}\tilde{\mathrm{S}}_{uu}^{\mathrm{b}}\\
%   %%
%          \tilde{\mathrm{S}}_{ud}^{\mathrm{a}} + \tilde{\mathrm{S}}_{uu}^{\mathrm{a}}\tilde{\mathrm{S}}_{ud}^{\mathrm{b}}\mathbb{D}_{dd}\tilde{\mathrm{S}}_{dd}^{\mathrm{a}}&
%          \tilde{\mathrm{S}}_{uu}^{\mathrm{a}}\mathbb{D}_{uu}\tilde{\mathrm{S}}_{uu}^{\mathrm{b}}
%   \end{pmatrix}.
 \end{equation}
% \editR{
% \begin{equation}
%     \tilde{\mathrm{S}}^{\mathrm{c}}=
%     \begin{pmatrix}
%     \tilde{\mathrm{S}}^{\mathrm{a}}_{ud} & 0\\
%     0 & \tilde{\mathrm{S}}^{\mathrm{b}}_{du}
%     \end{pmatrix}+
%     \begin{pmatrix}
% \tilde{\mathrm{S}}^{\mathrm{a}}_{uu} & 0\\
% 0 & \tilde{\mathrm{S}}^{\mathrm{b}}_{dd}
% \end{pmatrix} \begin{pmatrix}
% \tilde{\mathrm{S}}_{ud}^{\mathrm{b}}\mathbb{D}_{dd} & \mathbb{D}_{uu}\\
% \mathbb{D}_{dd} & \tilde{\mathrm{S}}_{du}^{\mathrm{a}}\mathbb{D}_{uu}
% \end{pmatrix}
% \begin{pmatrix}
% \tilde{\mathrm{S}}_{dd}^{\mathrm{a}} & 0\\
% 0 & \tilde{\mathrm{S}}_{uu}^{\mathrm{b}}
% \end{pmatrix}
% \end{equation}
% \begin{equation}
%     \begin{pmatrix}
% \tilde{\mathrm{S}}_{ud}^{\mathrm{b}}\mathbb{D}_{dd} & \mathbb{D}_{uu}\\
% \mathbb{D}_{dd} & \tilde{\mathrm{S}}_{du}^{\mathrm{a}}\mathbb{D}_{uu}
% \end{pmatrix}=\begin{pmatrix}
% \tilde{\mathrm{S}}_{ud}^{\mathrm{b}} & 1\\
% 1 & \tilde{\mathrm{S}}_{du}^{\mathrm{a}}
% \end{pmatrix}^{-1}
% \end{equation}
% }
\end{widetext}
\editR{where symbol $\mathbb{O}$ denotes the zero matrix.}

\begin{figure}[t!]
	\centering
	%\def\svgwidth{0.50\columnwidth} 
	%\includesvg{Figs/4}
	\includegraphics[width=1\columnwidth]{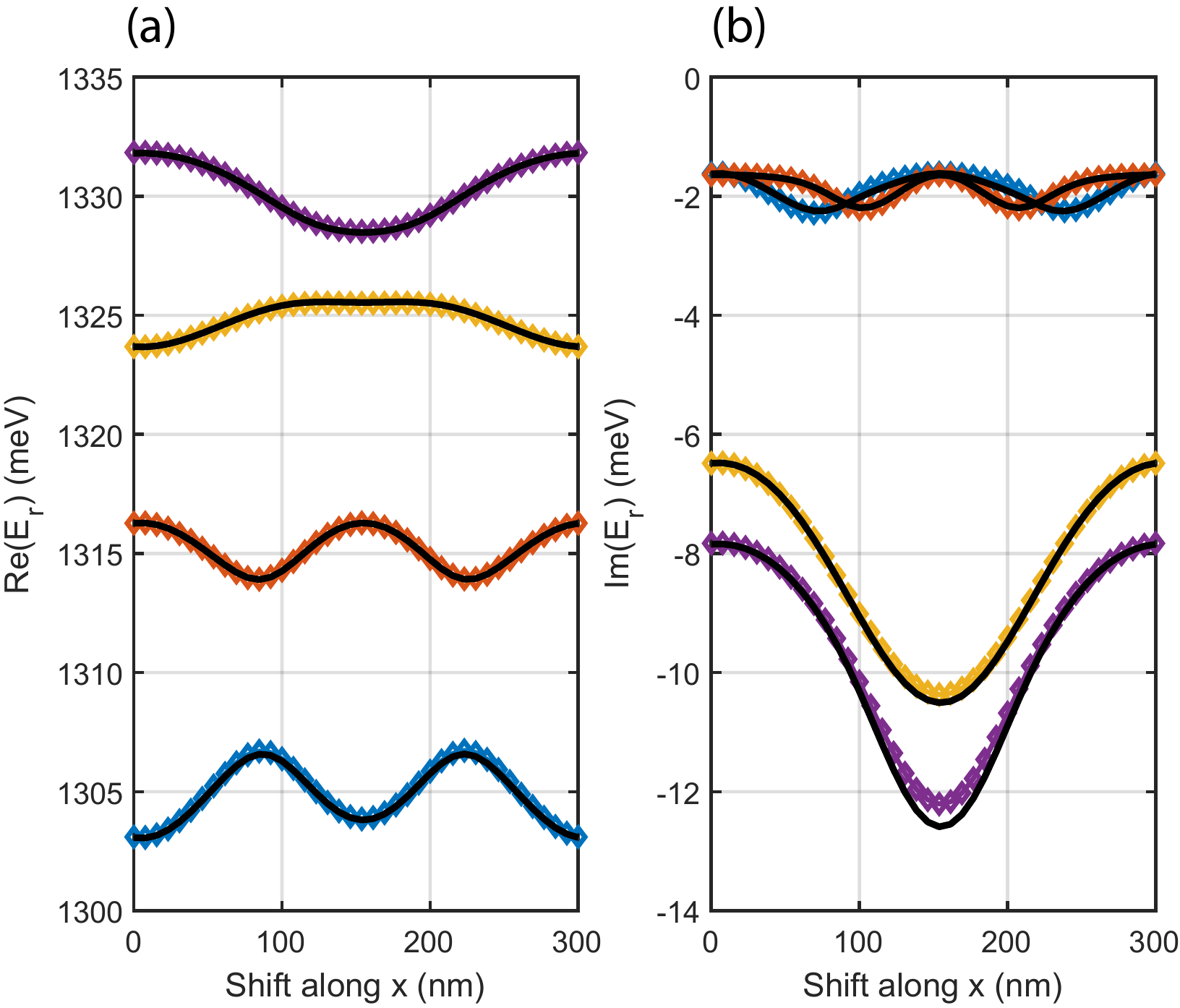}
	\caption{Panels (a,b): the real and imaginary parts of resonant energies of the stacked system calculated using the standard pole-search procedure (solid black line) and using the proposed here resonant mode coupling approximation (colored lines with diamonds). The background matrices are taken as constant and calculated at $\omega_{bg}=1325$~meV in the vicinity of one of the resonances.}
	\label{fig:ResEnergies}
\end{figure}
 {It appears that the resonant mode coupling approximation is very effective for modelling of the subsystems which can be shifted relatively to each other along the vertical z axis or the lateral axes x and y. In the first case, when subsystems A and B are separated by a homogeneous layer of nonzero thickness $h_g$ (see Fig.~\ref{fig:struct}), one can divide this layer in two pieces and consider them as parts of subsystems A and B. Another option is to introduce a propagation matrix $P_h$ as presented in Ref.~\cite{Weiss2011a}.}
%In the first case, when the subsystems A and B are separated by a homogeneous medium of thickness $L$, all equations remain the same except one should make the following substitutions:
% \begin{gather}
%     \left|O_{d}^{\mathrm{a}}\right\rangle\rightarrow \left|\hat{O}_{d}^{\mathrm{a}}\right\rangle=P\left|O_{d}^{\mathrm{a}}\right\rangle,\\
%     %%
%     \left|O_{u}^{\mathrm{b}}\right\rangle\rightarrow \left|\hat{O}_{u}^{\mathrm{b}}\right\rangle=P\left|O_{u}^{\mathrm{b}}\right\rangle,\\
%     %%
%     \tilde{\mathrm{S}}_{ud}^{\mathrm{b}}\rightarrow\hat{\mathrm{S}}_{ud}^{\mathrm{b}}=P\tilde{\mathrm{S}}_{ud}^{\mathrm{b}},\\
%     %%
%     \tilde{\mathrm{S}}_{du}^{\mathrm{a}}\rightarrow\hat{\mathrm{S}}_{du}^{\mathrm{a}}=P\tilde{\mathrm{S}}_{du}^{\mathrm{a}}
% \end{gather}
% where 
% \begin{equation}
%     P=\mathrm{diag}\{\exp(iK_zL)\}
% \end{equation}
% and $K_z$ is a vector consisting of z-components of the Fourier harmonics' wavevectors, calculated for the medium of the separation layer.
In the second case, when subsystem B is shifted relative to subsystem A by a distance $d_x$ along the x axis and by a distance $d_y$ along the y axis, then
the background matrix and the resonant input and output vectors of structure B should be transformed as:
\begin{equation}
\begin{aligned}
    \tilde{\mathrm{S}}^{\mathrm{b}}\rightarrow \tilde{\mathrm{S}}^{\mathrm{b'}}=& P^\dag\tilde{\mathrm{S}}^{\mathrm{b}}P,\;\\
    \left|O_n^{\mathrm{b}}\right\rangle\rightarrow\left|O_n^{\mathrm{b'}}\right\rangle= P^\dag\left|O_n^{\mathrm{b}}\right\rangle, \;&
    \left\langle I_n^{\mathrm{b}}\right|\rightarrow 
    \left\langle I_n^{\mathrm{b'}}\right|=\left\langle I_n^{\mathrm{b}}\right|P,
    \label{shifted}
    \end{aligned}
\end{equation}
where $P$ is a shift matrix defined as
\begin{equation}
    P=\mathrm{diag}\{\exp(iK_xd_x+iK_yd_y)\},
\end{equation}
with $K_{x,y}$ being the x and y projections of the wavevectors for the Fourier harmonics:
\begin{gather}
    K_x=k_x+G_x,\; K_y=k_y+G_y,\\
    \mathbf{G}=\{\dfrac{2\pi}{p_x}g_x,\dfrac{2\pi}{p_y}g_y,0\},\; g_{x,y}=0,\pm1,\pm 2,... \nonumber
\end{gather}
$k_{x,y}$ are the lateral projections of the incoming wave wavevector, $p_{x,y}$ are the structure periods, $\mathbf{G}$ is the reciprocal lattice.

Here it is important to discuss the applicability criteria of the described resonant mode coupling approximation.  The first one follows from the requirement %of unambiguously defined solution {(stability?)}: as we treat 
that we can treat the effective Hamiltonian in \eqref{Eigensystem} and \eqref{Eigen_EM} as an approximately constant matrix. At the same time, all the background matrices and as well as the Hamiltonian are, in fact, energy-dependent. This is why we require that all the background matrices are weak functions of energy in the selected energy range $[\omega_{min},\omega_{max}]$ used in the approximation: 
\begin{equation}
    %\tilde{\mathrm{S}}^{\mathrm{a, b}}\gg \dfrac{d\tilde{\mathrm{S}}^{\mathrm{a, b}}}{d\omega}(\omega_{max}-\omega_{min}).
    \tilde{\mathrm{S}}^{\mathrm{a, b}}(\omega\in[\omega_{min},\omega_{max}])\approx \mathrm{const}.
\end{equation}

The second applicability criteria is the absence of new Fabry-Perot resonances of the stacked system that could arise in matrices $\mathbb{D}_{uu,dd}$. Thus we require that in the approximation energy range:
\begin{equation}
    {\left|\delta\,\det\left(\mathbb{I}-\tilde{\mathrm{S}}_{u d}^{\mathrm{b}} \tilde{\mathrm{S}}_{d u}^{\mathrm{a}}\right)\right|}\ll{\left|\det\left(\mathbb{I}-\tilde{\mathrm{S}}_{u d}^{\mathrm{b}} \tilde{\mathrm{S}}_{d u}^{\mathrm{a}}\right)\right|},
\end{equation}
where we denote a function variation by $\delta$.

% One can try to construct a perturbation theory and include linear terms in energy
% \begin{equation}
% \tilde{\mathrm{S}}^{\mathrm{a, b}}(\omega)=\tilde{\mathrm{S}}^{\mathrm{a, b}}_0+\left.\dfrac{d\tilde{\mathrm{S}}^{\mathrm{a, b}}}{d\omega}\right|_{\omega=\omega_0}(\omega-\omega_0),
% \end{equation}
% but this will mean that all the resonant output and input vectors \eqref{inputOut} should be also calculated at the energy of the corresponding resonance. Further complication of the background matrices energy dependences will lead to the necessity to find poles of the nonlinear function 
% $(\omega-H(\omega))^{-1}$ so that the solution would be determined by equation \eqref{Eigen_EM} in the form:
% \begin{equation}
%     \begin{pmatrix}
%     \alpha\\
%     \beta
%     \end{pmatrix}=(\omega-H(\omega))^{-1}\left\langle I(\omega) \right|   \begin{pmatrix}
%     \left|d_1\right\rangle\\
%     \left|u_3\right\rangle
%     \end{pmatrix}.
% \end{equation}

\begin{figure*}
    \centering
    \includegraphics[width=1\linewidth]{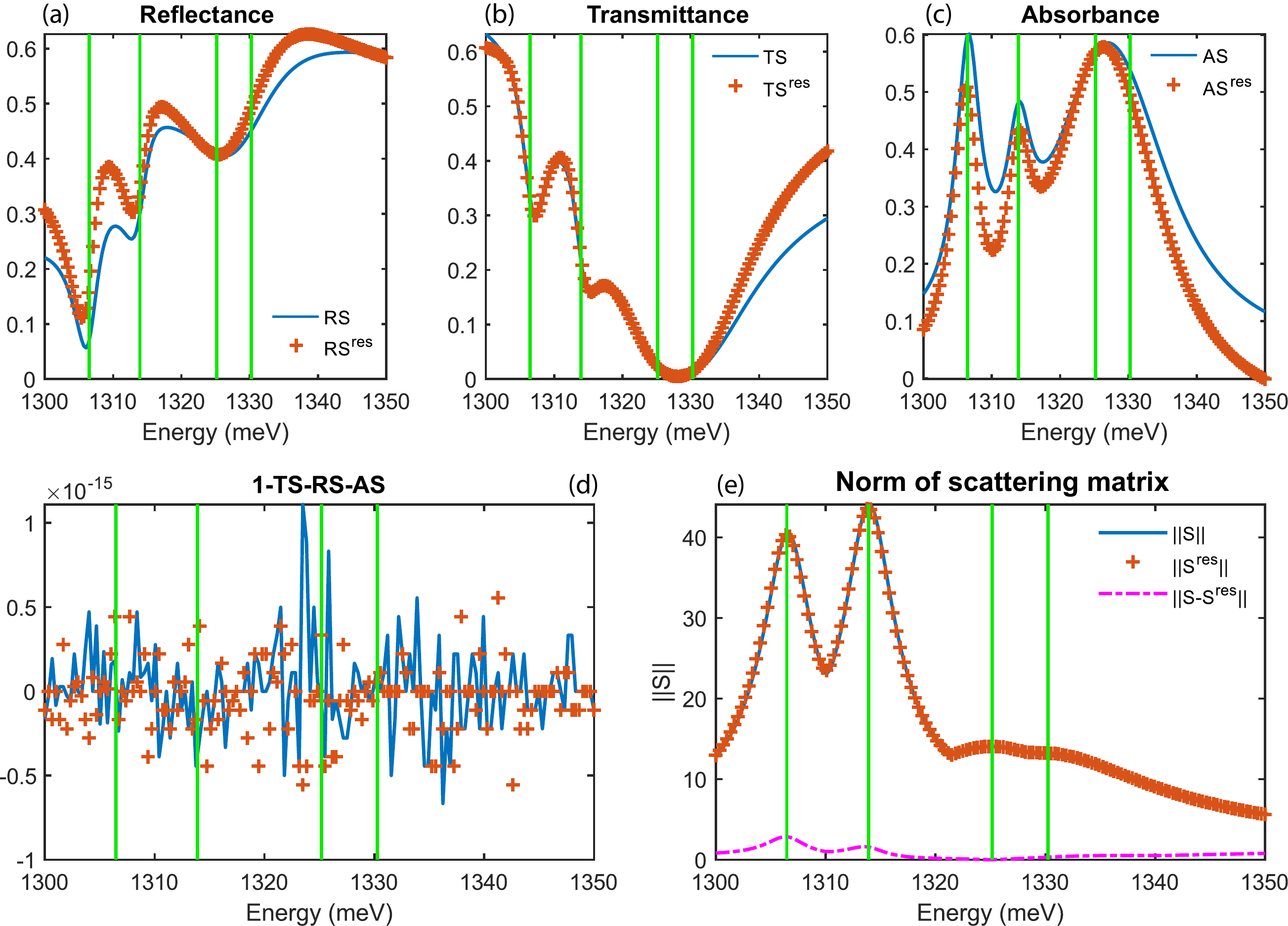}
    \caption{Energy dependence of the reflectance (a), transmittance (b), and absorbance (c) in S-polarization calculated by a standard FMM approach (blue line) and by the resonant mode coupling approximation (red crosses). (d) Deviation of the sum of these coefficients from unity. (e) Norm of the scattering matrix calculated by a standard FMM approach (blue line) and by the resonant mode coupling approximation (red crosses) as well as the norm of the difference between these two matrices. Vertical green lines denote the real part of the resonant energies. The calculations are conducted for the 90-nm shift of subsystem B along the x axis (see Fig.~\ref{fig:struct}). The background matrices are taken as constant and calculated at $\omega_{bg}=1325$~meV in the vicinity of one of the resonances. }
    \label{fig:Spectra}
\end{figure*}

\begin{figure*}
    \centering
    \includegraphics[width=1\linewidth]{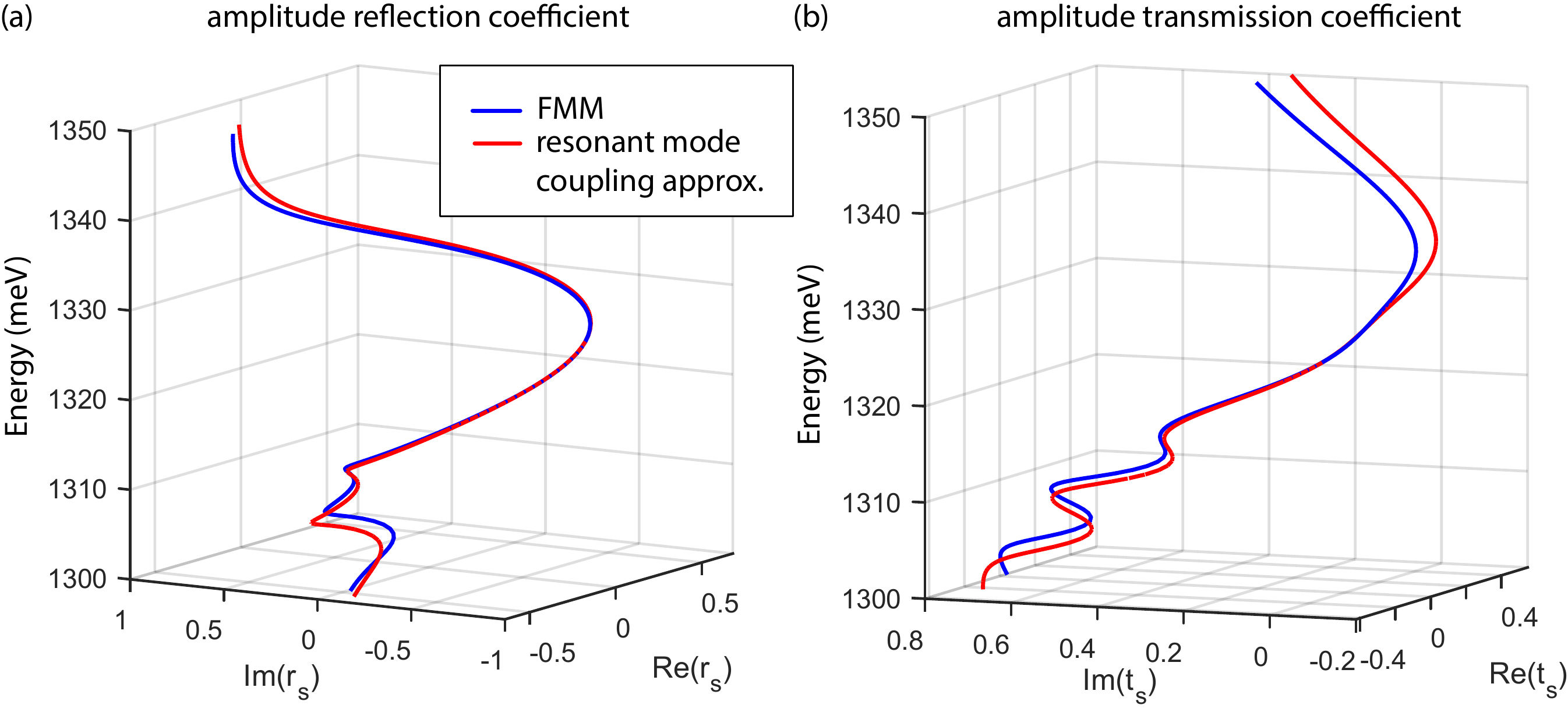}
    \caption{Energy dependence of the wave amplitude reflection (a) and transmission (b) coefficients in S-polarization calculated by a standard FMM approach (blue line) and by the resonant mode coupling approximation (red line). The calculations are conducted for the 90-nm shift of subsystem B along the x axis, $\omega_{bg}=1325$~meV. } %Vertical green lines denote the real part of the resonant energies.}
    \label{fig:3D}
\end{figure*}

\section{Numerical example}
\begin{figure*}
    \centering
    \begin{tikzpicture}
    \draw (0, 0) node[inner sep=0] {\includegraphics[width=0.8\linewidth]{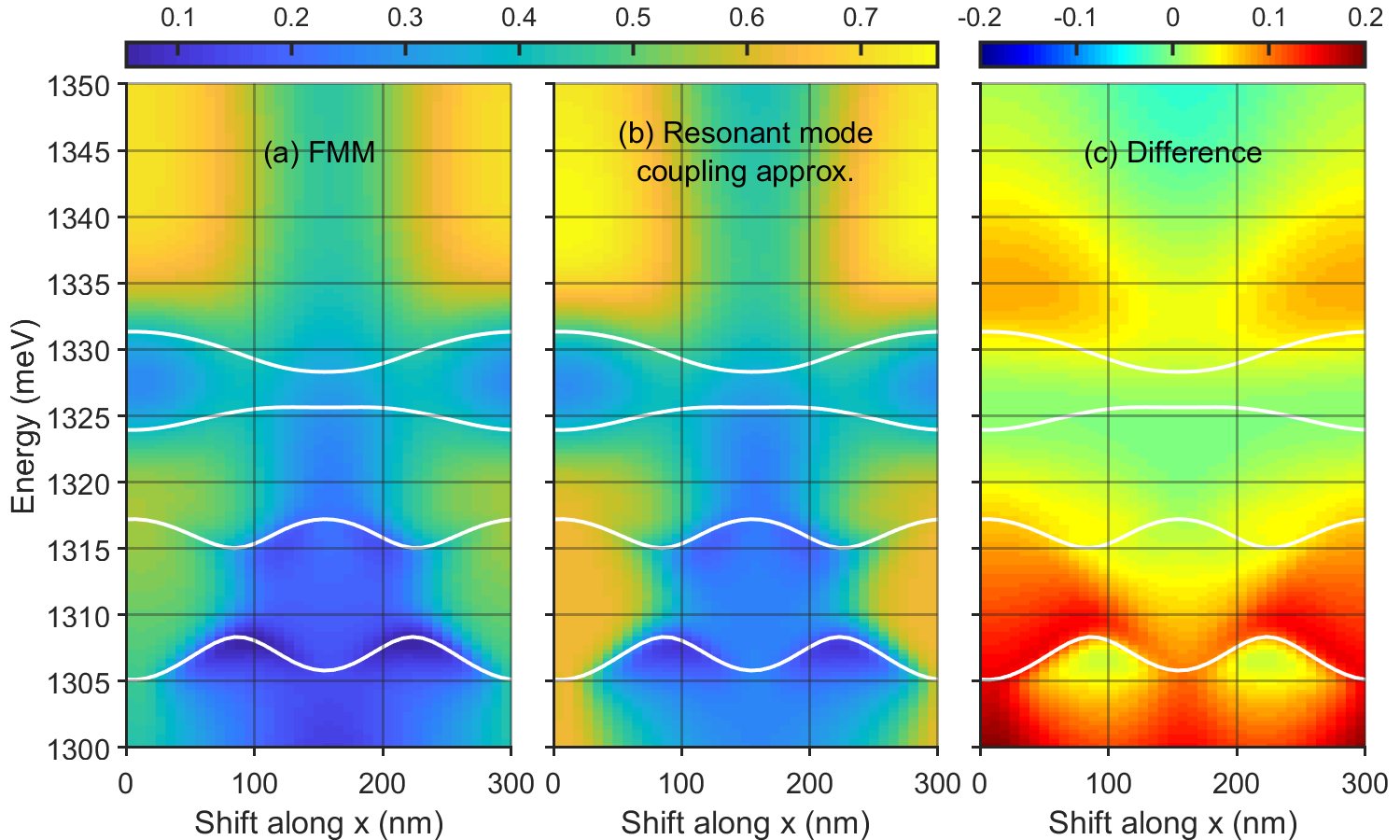}};
    %{1D_Grating_Bandstructure_H400_new.png}};
    % \draw (-0.3\linewidth, 0.24\linewidth) node[text width=4cm,align=center]%[anchor=west]
    %  {\textcolor{black}{\large{(a) FMM}}};
    %      \draw (0.2\linewidth, 0.24\linewidth) node[text width=4cm,align=center]%[anchor=west]
    %  {\textcolor{black}{\large{(b) Resonant mode \\ coupling approx.}}};
\end{tikzpicture}
    
    \caption{The reflectance spectra of S-polarized plane waves in dependence on the lateral shift of the structure B along the x axis calculated using (a) the FMM and (b) the resonant mode coupling approximation. Panel (c) presents the difference between the approximate and exact reflectances. The spectral positions of the resonances are illustrated with white lines. Background matrices of the subsystems are calculated at $\omega_{bg}=1325$~meV.}
    \label{fig:Surf}
\end{figure*}
For the numerical verification of the described approach, we conduct a series of optical spectra calculations for the structure that is sketched in Fig.~\ref{fig:struct}. Subsystem A consists of a 150-nm thick homogeneous crystalline Si layer \editR{with dielectric permittivity $\varepsilon_{Si}=13.03 + 0.033i$} and a 150-nm thick 1D periodic layer of SiO$_2$ strips \editR{with dielectric permittivity $\varepsilon_{SiO_2}=2.106$} embedded into the crystalline Si medium. The period of the structure is 300 nm, and the wires are 150-nm wide. The structure is periodic along the x axis and homogeneous along the y axis. The vertical axis z is directed downwards. Structure B is inverted structure A (the periodic layer on the top and the homogeneous on the bottom) with the only difference that the width of the SiO$_2$ strips is 160 nm. These two subsystems are surrounded by air; the distance between them is chosen to be infinitesimally small to provide maximum interaction of the coupled resonances via near fields. All following calculations are conducted for plane electromagnetic waves at normal incidence using the Fourier modal method with $N_g = 23$ Fourier harmonics along the periodicity axis. From now on, we use the convention that electromagnetic waves with the electric field directed along the y axis are called S-polarized waves.

Using the standard pole-search procedure \cite{Gippius2005c} we obtain two resonances for each subsystem $\omega^a_{1,2}=1309.43 - 1.47i$~meV, $1326.70-6.78i$~meV, $\omega^b_{1,2}=1316.02-1.49i$~meV, $ 1326.93-6.88i$~meV in the energy range $E=[1300,1350]$ meV. To derive the resonant energies of the stacked system in dependence on the lateral shift along the x axis we substitute the calculated resonant data into equation~\eqref{Eigensystem} with the proper shift operator \eqref{shifted}. One can compare the results obtained directly using the pole-search procedure for the scattering matrix of the whole system and the resonant energies calculated in the resonant mode coupling approximation (Fig.~\ref{fig:ResEnergies} (a)). %{The background scattering matrices $\tilde{\mathrm{S}}^{\mathrm{a,b}}$ are calculated at energy of 1325~meV. We treat them as constant matrices in the whole energy range of calculation.} %This evidentially fine accuracy was achieved through the use of an energy-dependent analytical expression for the background scattering matrices $\tilde{\mathrm{S}}^{\mathrm{a,b}}(\omega)=\tilde{\mathrm{S}}^{\mathrm{a,b}}_0+(\omega-\omega_0)\frac{d\tilde{\mathrm{S}}^{\mathrm{a,b}}}{d\omega}(\omega_0)$ calculated for $\omega_0=1325$ meV.    

Comparison of the FMM-calculated optical spectra and those derived in the resonant mode coupling approximation \eqref{inputOut} is presented in Fig.~\ref{fig:Spectra}. \editR{The reflectance and transmittance presented here are defined as squared modulus of the wave amplitude reflection and transmission coefficients, $R(\omega)=|r(\omega)|^2$ and $T(\omega)=|t(\omega)|^2$. The wave amplitude reflection and transmission coefficients are, in fact, complex-valued elements of the scattering matrix that correspond to the waves of zero diffraction order. Absorbance is defined as the part of the incoming energy flux that is neither transmitted, reflected, nor diffracted into any open channel.} Structure B is 90-nm shifted relative to structure A, exactly as depicted in Fig.~\ref{fig:struct}. The four calculated resonances of the stacked system allow us to reproduce the reflectance spectrum with proper accuracy. The background matrices of structures A and B are taken as constants and calculated at $\omega_{bg}=1325$~meV.

\editR{We find the results of the approximation to qualitatively reproduce all important features of the optical spectra. Indeed, all the Fano-shapes of the resonances and their spectral positions are preserved. Nevertheless, a concerned reader could wonder, what is the reason for the observed deviation from the FMM-calculated result. The 
accuracy of the proposed resonant mode coupling approximation
is governed by the accuracy of
the subsystems' background matrices approximation in the energy calculation range. 
In the proposed method the background matrices are calculated exactly at the fixed energy point $\omega_{bg}$. As a result, at $\omega=\omega_{bg}$ all spectral characteristics of the structure are computed to be exactly the same as in the case of direct FMM calculation (up to the machine precision).  At the same time, the almost perfect agreement between the approximated and exact scattering matrices (see Fig.~\ref{fig:Spectra}(e)) hints that the developed method must be more attractive in some other way. The correct way of testing the resonant mode coupling approximation is to compare the complex-valued amplitude reflection (or transmission) coefficients as functions of the photon energy. Such functions are represented by %ST non-flat 
curved lines in 3D space \{$\mathrm{Re}(r)$, $\mathrm{Im}(r)$, $\omega$\} and \{$\mathrm{Re}(t)$, $\mathrm{Im}(t)$, $\omega$\} (see Fig.~\ref{fig:3D}). %; the squared absolute value of this function at each energy $\omega$  gives what we usually call a reflectance (or transmittance). 
As shown in Fig.~\ref{fig:3D}, the approximated and exact lines are very close to each other, and the resonant mode coupling approximation describes correctly all the bends and loops of this trajectory. It might happen that at some energy, the exact trajectory comes to zero reflection (or transmission) much closer than the approximated one. This situation leads to a high relative error of the approximation method in terms of the reflectance $R$ or transmittance $T$, although nothing special occurs at this energy on the complex 3D diagram. Remarkably, the same degree of agreement is reminiscent of comparing the experimental and theoretical curves, where at some energies the relative difference can be fairly large. }

In order to provide a more comprehensive demonstration of the method's applicability, we calculate the reflectance spectra for the structures with 40 shift values ranging from $d_x=0$ to $d_x=p$. Results of the exact FMM calculation and the resonant mode coupling approximation calculation are presented in Fig.~\ref{fig:Surf}. The maximum difference between the exact and approximate reflectances is 0.195. \editR{We consider this accuracy to be good enough to reproduce the main features of the nanostructures' optical properties. Moreover, we observe more than 200-fold calculation acceleration for the structure of interest with the parameters listed above, and an energy grid of 150 points (assuming that the resonant vectors and energies of subsystems A and B are already found). This acceleration grows proportionally to the number of layers in the subsystems, it also increases with the number of harmonics included.}  %: the calculation time on a personal computer of the data presented in Fig.~\ref{fig:Surf} with the exact method is 105 seconds vs. 0.5 seconds with the resonant mode coupling approximation provided that the resonances of the subsystems are known.
\editR{
\section{Discussion}

    Comparing the presented extension of the resonant mode coupling approximation with the previous works Refs.~\cite{Gippius2010,Weiss2011a} we should underline the following significant improvements. With the possibility to calculate a full scattering matrix, the proposed method can now be iteratively applied to any number of subsystems. Given two subsystems characterized by resonant energies, input, and output vectors, the calculation outcome for the stacked system is of the same structure and can be used for further recursive calculations.
This allows us, unlike within the previously published resonant approximations, to calculate almost instantaneously the scattering matrices of a continuous range of systems composed of differently combined subsystems provided that the resonant energies, input, and output vectors of subsystems are known. Additionally, not
only can we determine the resonant energies, but we can also predict the Fano-shape of each resonance (i.e., determine whether this resonance corresponds to a dip or a peak in a reflectance or transmittance spectrum). In most cases of the experiment to theory comparisons, this information appears to be physically sufficient as experimentally measured optical spectra of a fabricated structure usually undergo blurring and broadening due to the structure imperfections and experimental setup peculiarities. 

Still, the major drawback of the method is calculation inaccuracy that happens due to poor approximation of the background scattering matrix. One can see that the compared optical spectra differ the most on the boundaries of the investigated energy range, while at the energy 1325~meV of the background matrix calculation, they are identical. Thus, further development of the resonant coupling method should be focused on the better approximation of the background matrix. 

The proposed method provides an analytical and rigorous description of the system under consideration. Coupling between the resonant modes could be derived without employing any phenomenological parameters. Such analytical representation allows direct investigation of the interaction of the resonant modes. Moreover, output resonant vectors and intermediate amplitudes vectors $\left|d_2,u_2\right\rangle$ \eqref{eq_d2}, \eqref{eq_u2} could be used for the calculation of the resonant mode fields distribution. While the Fourier modal method is a purely computational tool that constructs the solution of the linear electromagnetic problem as a sum over plane waves, the resonant mode coupling approximation introduces a clear physical meaning due to explicit usage of the system resonances.

The modeling structure that we consider here is a toy example representing all necessary features of an arbitrary stacked system. While for 1D periodic system the number of harmonics $N_g$ we chose is relatively low, using 20-30 harmonics along each axis (400-900 total) is a standard practice for the 2D periodicity. As the calculation time for the Fourier modal method grows approximately as $N_g^{2.3}$ (or $N_g^3$ if we do utilize straightforward algorithms of matrix multiplication), the proposed coupled resonances approximation would provide a substantial speedup for 2D-periodic structures. Here we should highlight that the provided formulation of the method is universal and could be applied to systems with 2D periodicity, 1D periodicity or no periodicity at all.
}
\section{Conclusion}
In conclusion, we propose a further development of the resonant mode coupling approximation method of Refs.~\cite{Gippius2010, Weiss2011a}. We consistently derived the input and output resonant vectors of the stacked system, upper and lower parts of which exhibit distinctive resonances. We show that the new method can reproduce the exact calculation results with reasonable precision. \editR{Effective Hamiltonian formulation reveals the physical nature of the resonant coupling, while explicit formulas for the input and output resonances allow one to use the calculation technique recursively.} At the same time, the proposed method dramatically decreases the computational time and opens new possibilities for using resonances of nanostructures not only in the analysis of optical spectra but also as a powerful optimization tool. For example, one can prepare a database of optical resonances in simple photonic structures and use the resonant mode coupling approximation to calculate the optical spectra of stacked systems. This calculation procedure could be used in machine learning and genetic algorithms to find the best structure design for enhancing a selected optical property. The developed approach can be implemented even on personal computers since each resonance is described by two resonant vectors and their complex energy. 
\acknowledgements{This work is funded by the Russian Foundation for Basic Research (Grant No. 18-29-20032).}
 %\bibliography{JAB_library}
 %\acknowledgements
 %
\end{document}